\documentclass[%
 reprint,
 amsmath,amssymb,
 aps,
]{revtex4-1}

\usepackage{graphicx}% Include figure files
\usepackage{dcolumn}% Align table columns on decimal point
\usepackage{bm}% bold math
\usepackage{amssymb}
\usepackage{xspace}

\usepackage{hyperref}
\usepackage{amsmath}

\usepackage{color}

\usepackage{ulem}

\begin{document}
\newcommand{\pp}{\ensuremath{\rm pp}\xspace}
\newcommand{\St}{\ensuremath{\rm S_{T}}\xspace}
\newcommand{\Raa}{\ensuremath{\rm R_{AA}}\xspace}
\newcommand{\ppb}{\ensuremath{\rm p\!-\!Pb}\xspace}
\newcommand{\cme}{\ensuremath{\sqrt{s}}\xspace}
\newcommand{\pt}{\ensuremath{p_{\rm{T}}}\xspace}
\newcommand{\st}{\ensuremath{S_{\rm{T}}}\xspace}
\newcommand{\so}{\ensuremath{S_{\rm{0}}}\xspace}

\newcommand{\gev}{\ensuremath{{\rm GeV}/c}\xspace}
\newcommand{\pbpb}{\ensuremath{\rm Pb\!-\!Pb}\xspace}
\newcommand{\auau}{\ensuremath{\rm Au\!-\!Au}\xspace}
\newcommand{\ptopi}{\ensuremath{{\rm p } / \pi}\xspace}
\newcommand{\ktopi}{\ensuremath{({\rm K}^{+}+{\rm K}^{-}) / (\pi^{+}+\pi^{-})}\xspace}
\newcommand{\twotwo}{\ensuremath{2\rightarrow 2}\xspace}
\newcommand{\meanpt}{\ensuremath{\langle p_{\rm T} \rangle}\xspace}
\newcommand{\ltok}{\ensuremath{({\rm \Lambda}^{0}+\bar {\rm \Lambda}^{0})/(\rm 2 K^{0}_{s} )} \xspace}
\newcommand{\phitopi}{\ensuremath{(\rm 2 \phi)  / (\pi^{+}+\pi^{-})} \xspace}
\newcommand{\added}[1]{\textcolor{red}{#1}}
\newcommand{\removed}[1]{\sout{#1}}

\preprint{AIP/123-QED}

%\title{Spheri(o)city approach to study the soft and hard contributions \\ to the identified hadron \pt spectra}% Force line breaks with \\

\title{Disentangling the soft and hard components of the \\ pp  collisions  using the sphero(i)city approach}% Force line breaks with \\

\author{E. Cuautle Flores, R. T. Jimenez Bustamante,  I. A. Maldonado Cervantes, \\ A. Ortiz Velasquez, G. Pai\'c and E. P\'erez Lezama}
% \altaffiliation{mail }%Lines break automatically or can be forced with \\
%\author{G. Pai\'c}%
%\author{Tonatiuh}%
%\author{Edgar}%
%\author{Antonio}%

\affiliation{%
 Instituto de Ciencias Nucleares, Universidad Nacional Aut\'onoma de M\'exico \\
 Apartado Postal 70-543, M\'exico Distrito Federal 04510, M\'exico. %\textbackslash\textbackslash
}%

\date{\today}% It is always \today, today,

\begin{abstract}

A new method to extract information from the
\pp data is proposed. The approach is based on the use of the event structure variables:
sphericity and spherocity, to split the data into enhanced soft and hard processes samples corresponding to events with large and low numbers of multi-parton interactions, respectively. The present study was developed in the framework of Pythia 8.180 for inelastic \pp collisions at $\sqrt{s}=$7 TeV. As an application of the method, a study of the identified particle transverse momentum spectra and their ratios; is presented for soft (isotropic) and hard (jetty-like) events. The flow-like effect on these observables due to multi-parton interactions and color reconnection is relevant for soft events suggesting that partons inside the jet do not feel color reconnection and its flow-like consequences.

% Due to the flow-like effect incorporated in Pythia which increases with the increasing number of multi-partonic interactions, the baryon-to-meson ratio increases with increasing spheri(o)city. Other ratios like $\phi/\pi$ are also larger for isotropic than for jetty-like events. The implementation of this kind of analysis in LHC data could help to improve their description by models and also help to understand the recent results in heavy ion collisions.

\end{abstract}
                          
\pacs{13.85.Hd,13.85.Ni,13.87.Fh}% PACS, 
%\keywords{jets, hard and soft processes} % keywords

\maketitle  
 %%%%%%%%%%%%%%%%%%%%%%%%%%%%%%%%%%%%%%%%%%
%\section{Introduction}  
%%%%%%%%%%%%%%%%%%%%%%%%%%%%%%%%%%%%%%%%%% 

Due to the composite nature of hadrons, it is possible to have  events
in which two or more distinct hard parton-parton interactions occur
simultaneously in a single hadron-hadron collision.  The phenomenon, named multi-parton interactions (MPI)~\cite{PhysRevD.36.2019}, has been
supported by  data~\cite{Chekanov:2007ab,Abe:1997bp,Abazov:2009gc} and is a key
ingredient in the Monte Carlo event
generators~\cite{Sjostrand:2006za}. Its parameters are typically tuned
using global observables like multiplicity, transverse momentum (\pt)
and their correlations. However, this approach leaves out many of the
details of the \pp interactions  making hard to describe  observables like the jet production rate
as a function of the event multiplicity ~\cite{Abelev:2012sk,Chatrchyan:2013ala,Aad:2012fza,Aad:2012fza}, the strange particle production~\cite{Abelev:2012hy,Abelev:2012jp,Aamodt:2011zza} and the approximately linear increase of the J$/\psi$
yield with the event multiplicity~\cite{Abelev:2012rz}. New observables are needed to understand which component of the hadronic interactions, hard (pQCD) and/or soft (phenomenological models), is worst described by theory causing the overall disagreements.

In this letter observables which usually are not considered in the analysis of minimum bias data are shown to bring a new insight into the fine structure of the events. In an earlier letter~\cite{Ortiz:2013yxa} it was demonstrated that the color reconnection (CR) mechanism implemented in Pythia 8.180~\cite{Sjostrand:2007gs} creates flow-like patterns in \pp collisions which increases with the number of MPI (nMPI). The average nMPI rises almost proportionally with the event multiplicity, but high multiplicity events are affected by  a fragmentation bias~\cite{Abelev:2013sqa}.  The present work  goes a step further demonstrating the possibility to   separate special low and high nMPI events to isolate  the behavior of particles inside  jets and those pertaining to the bulk (soft processes) - the low and high spheri(o)city, respectively.  A similar treatment may be applied to heavy ion collisions where  the 
enhancement of the baryon-to-meson ratio with respect to \pp
collisions is not fully understood~\cite{Abelev:2014laa}.

%
%Although nMPI rises almost proportionally with the event multiplicity, the latter can not be used to chose events with high flow-like due to a fragmentation bias~\cite{Abelev:2013sqa}.  Instead, the present work  demonstrates the possibility to   separate special high and low nMPI  to isolate  the behavior of particles inside  jets and those pertaining to the bulk (soft processes) - the low and high spheri(o)city, respectively.  A similar treatment may be applied to heavy ion collisions where  the 
%enhancement of the baryon-to-meson ratio with respect to \pp
%collisions is not fully understood in terms of medium effects~\cite{Abelev:2014laa}.

 \begin{figure}[htbp]%[t] %[htbp]
\begin{center}
   \includegraphics[width=0.5\textwidth]{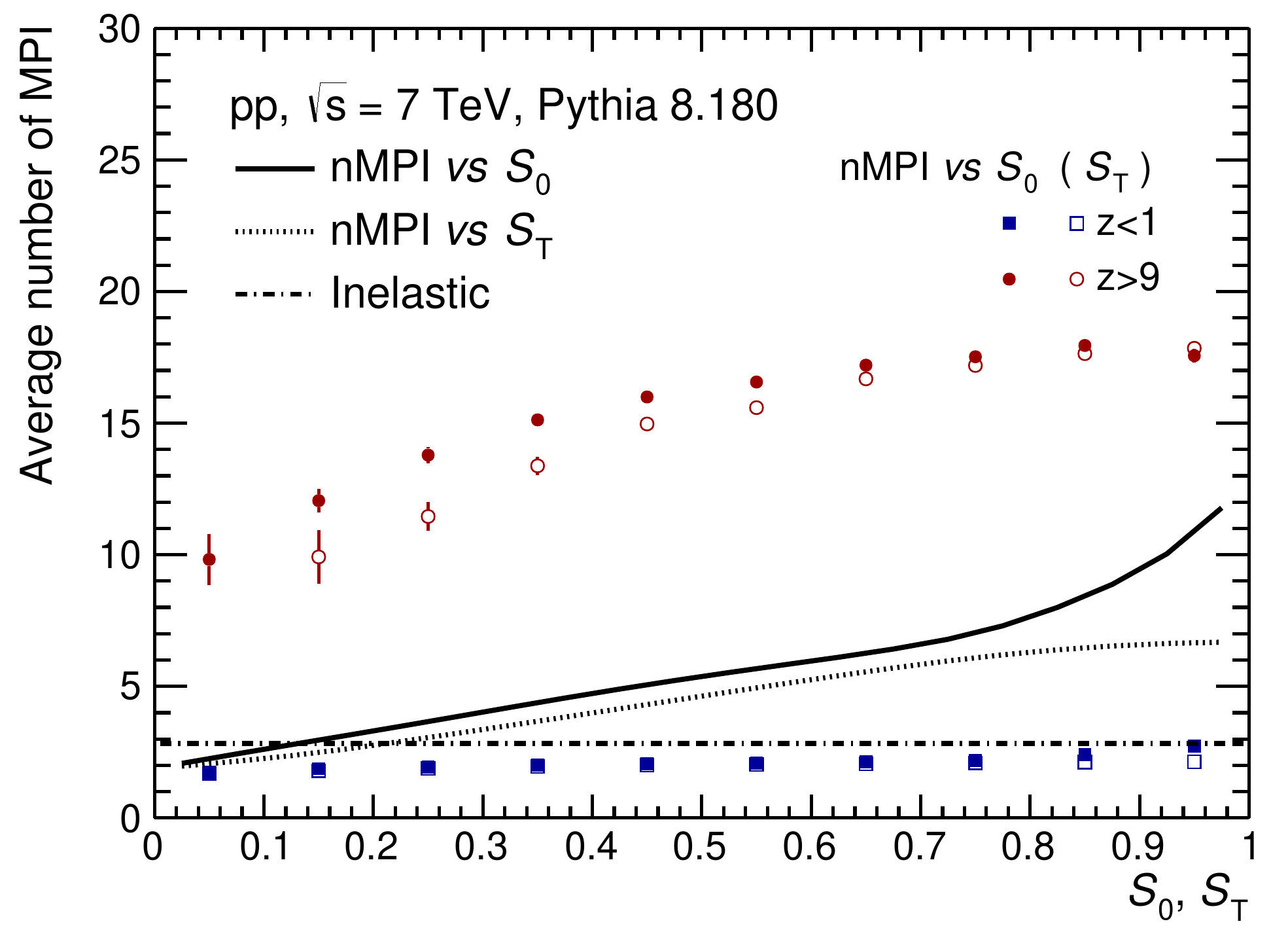}
   \caption{(Color online) Correlation between the number of multi-partonic interactions, nMPI, and sphero(i)city. The solid and dotted lines indicate the the average nMPI as a function of spherocity and sphericity, respectively. The horizontal line shows the value for inelastic \pp collisions. Also results for low and high multiplicity, $z$, are shown.}
  \label{Fig1}
\end{center}
\end{figure}

%%%%%%%%%%%%%%%%%%%%%%%%%%%%%%%%%%%%%%%%%%%%%%%%%%%%%%%%%%%%%%%%%%%%%%%%%
%\section{Event Shape variables}
%%%%%%%%%%%%%%%%%%%%%%%%%%%%%%%%%%%%%%%%%%%%%%%%%%%%%%%%%%%%%%%%%%%%%%%%%

\begin{figure}[htbp]%[t] %[htbp]
\begin{center}
   \includegraphics[width=0.5\textwidth]{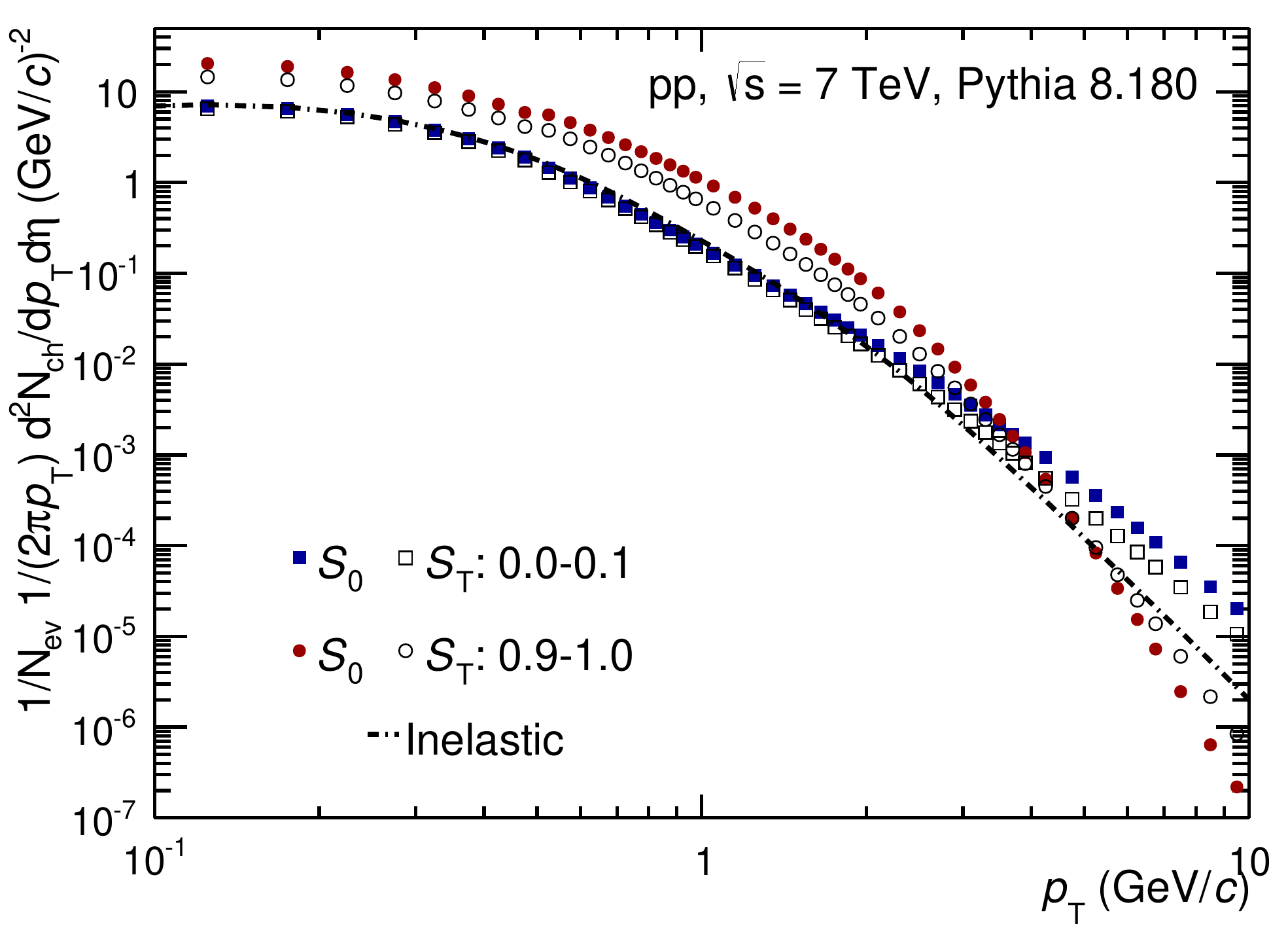}
   \caption{ (Color online) Transverse momentum distributions of
     jetty-like and soft events selected either with sphericity
     (empty markers) or spherocity (full markers). A comparison with
     the inclusive distribution (line) is also shown. Distributions
     were calculated for primary charged particles in $|\eta|<1$.}
  \label{Fig2}
\end{center}
\end{figure}

At hadron colliders, the event shape variables measure the geometrical
distribution of the \pt's of the outgoing hadrons. The restriction to
the transverse plane avoids  the  bias   from  the  boost   along  the
beam axis~\cite{Banfi:2010xy}. The variables transverse
sphericity ($S_{\rm T}$) and spherocity ($S_{\rm 0}$) are used in this
work, by definition both are collinear safe.  $S_{\rm T}$ is
defined in terms of the eigen-values $\lambda_{1}$, $\lambda_{2}$
($\lambda_{1}>\lambda_{2}$):

\begin{equation}
S_{\rm T} = \frac{2\lambda_{2}}{\lambda_{2}+\lambda_{1}}\,,
\end{equation}
resulting from the diagonalization of the transverse  momentum matrix:

\begin{displaymath}
\mathbf{S_{xy}} = \frac{1}{\sum_{j} {p_{\rm T}}_{j}}\sum_{i}
	\frac{1}{ {p_{\rm T}}_{i} }\left(\begin{array}{cc}
{p_{\rm x}}_{i}^{2}      &  {p_{\rm x}}_{i} \, {p_{\rm y}}_{i} \\
{p_{\rm y}}_{i} \, {p_{\rm x}}_{i} &  {p_{\rm y}}_{i}^{2} 
	\end{array} \right).        
\label{TSph}
\end{displaymath}

Transverse spherocity is defined for a unit transverse vector $\mathbf{\hat{n}}$ which minimizes the ratio below:

\begin{equation}
S_{\rm 0} = \frac{\pi^{2}}{4} \left(  \frac{\sum_{i} {{\overrightarrow{\pt}}}_{i} \times \mathbf{\hat{n}}}{\sum_{i} {\pt}_{i}}  \right)^{2}.
\end{equation}

By construction, the limits of the variables are related to specific configurations in the transverse plane
\begin{displaymath}
%S_{\rm T}=\left(\lbrace  \begin{array}{ll}
S_{\rm T}  (S_{\rm 0}) = \left\lbrace    \begin{array}{ll}
0 & \textrm{``pencil-like'' limit (hard events)} \\
1 & \textrm{``isotropic'' limit (soft events)}
\end{array} \right. \,.
\end{displaymath}

The event shape variables were studied for inelastic \pp collisions generated
with Pythia 8.180 tune 4C~\cite{Corke:2010yf}, about $2.5\times10^{9}$
events were produced for each set of simulations. More than two
primary charged particles having \pt above 500 MeV/$c$ in $|\eta|<0.8$
were requested for each event to ensure that experiments at the LHC be
able to perform this kind of analysis. Only $\approx$49\% of the
events satisfy these requirements. The jetty-like events discussed
here are those having $S_{\rm 0}$ ($S_{\rm T}$) below 0.1 and
represent 3.51\% (1.84\%) of the inelastic cross section, while the
soft ones have event shape values above 0.9 and correspond to
0.06\% (1.72\%) of the inelastic cross section.

%%%%%%%%%%%%%%%%%%%%%%%%%%%%%%%%%%%%%%%%%%%%%%%%%%%%%%%%%%%%%%%%%%%%%%%%%%%%%%%%%%%%%%%
%\section{Results}
%%%%%%%%%%%%%%%%%%%%%%%%%%%%%%%%%%%%%%%%%%%%%%%%%%%%%%%%%%%%%%%%%%%%%%%%%%%%%%%%%%%%%%%

 \begin{figure*}[htbp]%[t] %[htbp]
\begin{center}
   \includegraphics[width=0.99\textwidth]{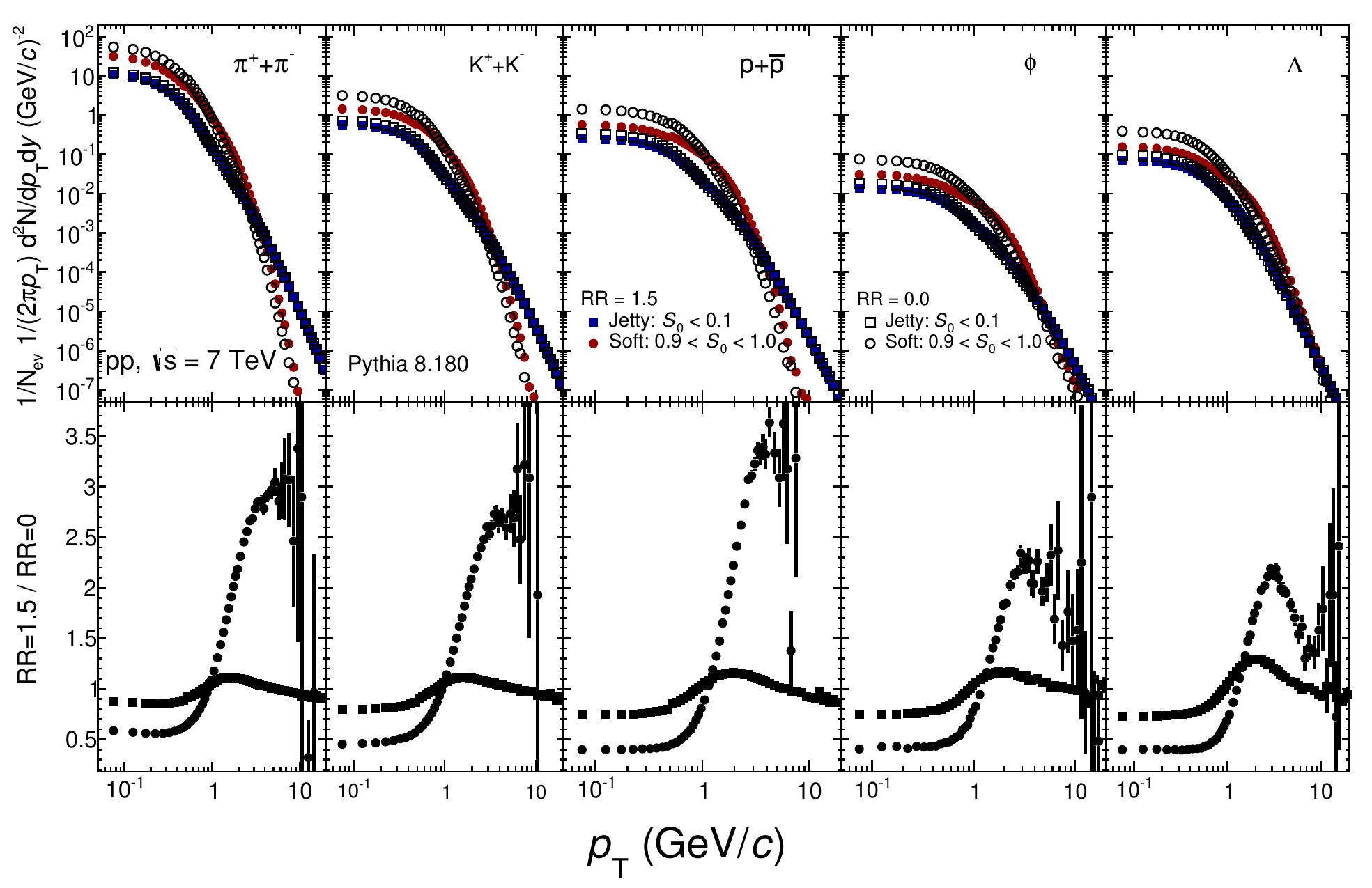}
   \caption{ (Color online) Transverse momentum spectra of
     $\pi^{\pm}$, ${\rm K}^{\pm}$, ${\rm p(\bar{p})}$, $\phi$ and
     $\Lambda$ obtained for \pp collisions at $\sqrt{s}=7$ TeV
     simulated with Pythia 8.180 tune 4C are shown in the top panel. Results for jetty-like (squares) and
     soft (circles) events are compared. To illustrate the 
     flow-like effect on the spectra~\citep{Ortiz:2013yxa}, two event
     classes simulated with (full markers) and without (empty
     markers) color reconnection are presented. Spectra were measured
     for primary charged particles in $|y|<0.5$. The ratio with to without color reconnection is shown in the bottom panel. }

  \label{Fig3}
\end{center}
\end{figure*}

%The correlation between event shapes and quantities at the partonic
%stage was studied. Figure~\ref{Fig1} shows nMPI as a function of sphericity and spherocity. The average nMPI increases going from the pencil-like limit to the isotropic
%limit. This makes the tool so important
%for selecting and studying events with large flow-like
%effects~\citep{Ortiz:2013yxa}. On the contrary, the jetty-like events come
%from interactions where the average nMPI is below that for the
%inclusive sample ($\approx 2.7$). Hence, in those events the effect of the underlying
%event (everything except the jet) is expected to be significantly
%suppressed allowing the study of (mini) jet properties. It is worth noting that the average nMPI is larger when the selection is done in
%\so instead of \st, for instance, close to the isotropic limit it
%gives a $\approx$71\% enhancement on nMPI when selection is done in \so with respect to \st.

The correlation between event shapes and quantities at the partonic
stage was studied. Figure~\ref{Fig1} shows  several curves corresponding to the average nMPI as a function of \st and \so. Results are shown for the entire sample (inclusive) and for two extreme values of multiplicity given in terms of the variable $z=({\rm d}N/{\rm d}\eta)/\langle {\rm d}N/{\rm d}\eta \rangle$. The main features of the correlation plots are the following: i) nMPI increases proportionally with the value of \so or \st. For the inclusive sample, \so nMPI  increases sharply at high \so values. ii) The dependence for high $z$ saturates around 17 in the high \so and \st ranges. iii) The distributions  for \st and \so at high $z$ are more similar than for the inclusive ones.

Spherocity is more effective in discriminating multi-jet topologies
than sphericity~\cite{Banfi:2010xy}. The effect of this feature can be
seen in Fig.~\ref{Fig2} where the transverse momentum distributions of
primary charged particles are shown for soft and jetty-like
events selected either with \so or \st. The production of low \pt
particles ($\pt<3$ GeV/$c$) for soft events is significantly
larger than in the inclusive sample, while, the production
of particle with larger \pt is reduced. For jetty-like events, the
particle production for $\pt<2$ GeV/$c$ is similar to that 
of the inclusive sample, for larger \pt the power law tail
is more prominent than that observed in inelastic events. Finally, for soft events a small power law tail is visible; stronger for \st than for \so in line with the existence of multijet events at large \st. 

%
%
%Finally, for soft events a stronger power law tail is observed when selection
%is done in \st instead \so. This is consistent with a larger multi-jet
%contribution in high \st events.

In the following, only results using spherocity are presented since with this
observable the largest nMPIs events can be selected. The top panel of Fig.~\ref{Fig3} shows the invariant yields, ${\rm
  d}N/{\rm d}y$, of primary $\pi^{+}+\pi^{-}$, $\rm{K^{+}+K^{-}}$, $\rm{p+\bar{p}}$, $\phi$-meson and $\Lambda$-baryon for low and
high \so events computed for $|y|<0.5$. Results for two sets of simulations are shown, the first one
including color reconnection using the  value for the
reconnection range, RR$=1.5$, as in tune 4C. The second simulation uses RR$=0$ (without CR). This comparison is motivated by the fact that CR produces flow-like effects which
are augmented in events with high nMPI. Color reconnection makes the spectra harder in a \pt region around $2$
GeV/$c$, the hardening increases with the hadron mass. This effect is not observed neither in high \so events when RR$=0$ nor
in jetty-like events (with and without CR). Soft events are more
affected by CR than jetty-like ones due to the different average nMPI. To quantify the effects on both
sub-samples the yields obtained with CR are normalized to those where
RR$=0$, the ratios are plotted in the bottom panel of
Fig.~\ref{Fig3}.   The ratios for $\pi/{\rm K/p}$ are close to one (within 10\%) for jetty like events, while, for high \so events they are significantly higher. Indeed, this
ratio shows up a maximum which is reached at higher \pt
when RR is not zero. It is worth to note that the mass effect is
small for jetty-like events, {\it e.g.,} the functional forms of the ratios as a function of \pt for pions
and protons are almost equal, but for high \so events this
is not the case.

\begin{figure*}[htbp]%[t] %[htbp]
\begin{center}
   \includegraphics[width=0.99\textwidth]{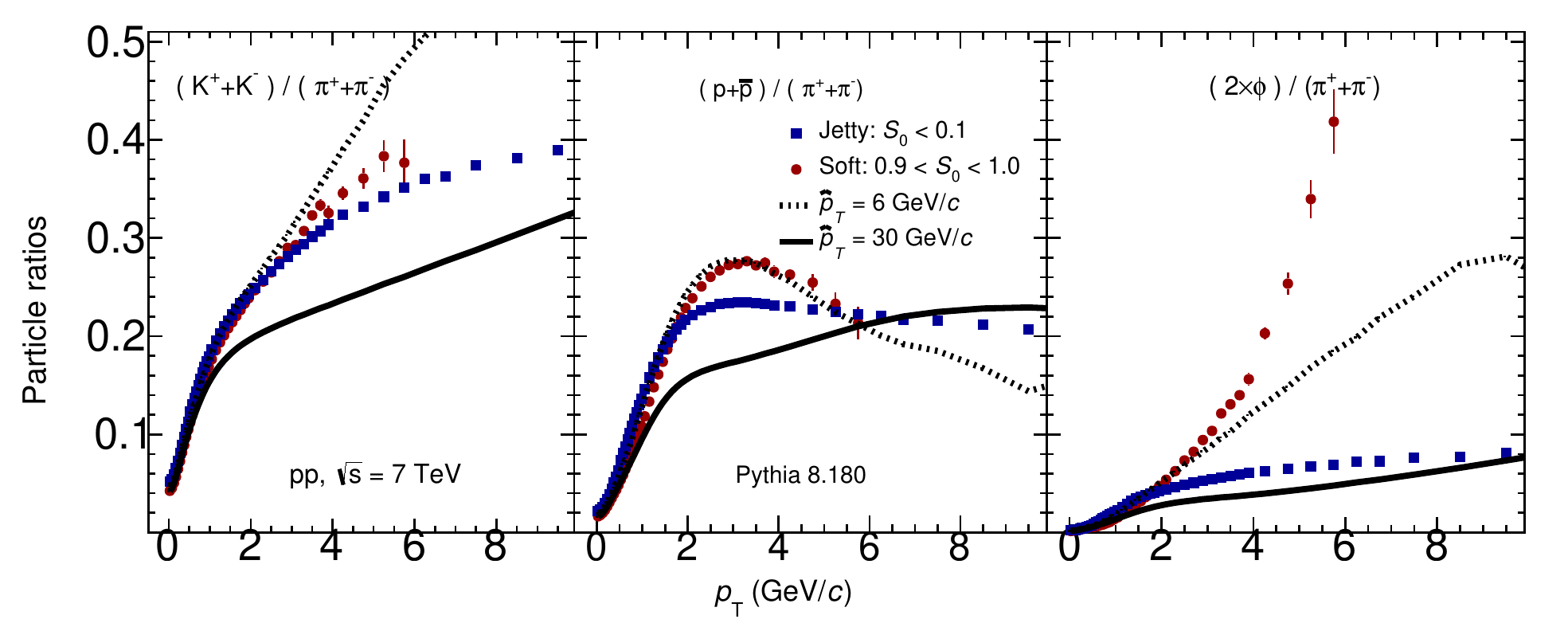}
   \caption{ (Color online) Particle ratios as a function of \pt for
     jetty-like (squares) and soft events (circles) obtained in
     inelastic \pp collisions simulated with Pythia 8.180 tune
     4C. Results are compared to \pp collisions where the partonic
     transverse momentum of the most energetic interaction,
     $\hat{p}_{\rm T}$, was set to 6 (dotted line) and 30 (solid line)
     GeV/$c$.}
  \label{Fig5}
\end{center}
\end{figure*}

Figure~\ref{Fig5} shows the particle ratios, {\it i.e.,}  the yields,
${\rm d}N/{\rm d}y$, of $\rm{K^{+}+K^{-}}$, $\rm{p+\bar{p}}$, $\phi$-meson normalized to
that for pions. The results are presented for jetty-like and soft
events, the simulations included CR. For \pt below 2 GeV/$c$ the
ratios exhibit a depletion going from low to high \so, for
larger \pt the ratios are higher in high \so events. For \pt above
5-6 GeV/$c$ the proton-to-pion ratio in high \so events returns to the
value obtained for low \so. The other two ratios which involve
strange hadrons deviate from each other when the transverse momentum
exceeds 2 GeV/$c$. These results are compared to those where the
inelastic \pp interactions have a partonic transverse momentum, $\hat{p}_{\rm T}$, of 6 or 30 GeV/$c$, being
this the most energetic pQCD process. The sample with 6 GeV/$c$ jets is in
a qualitatively good agreement with results for spherical events, this is due to the fact that the particles inside jets with  $\hat{p}_{\rm T}=$ 6 GeV/$c$ result in jet of very small energy likely to be in the realm of soft processes. As
the energy of the jet increases the corresponding particle ratios decrease. This
feature is also seen when the results  for soft events are compared to those for the hard ones using the spherocity approach. In
Table~\ref{tab:table1} the \pt integrated particle ratios are reported in spherocity
intervals. Within $\approx$10\% they are found to be the same in the soft and hard interactions. This suggests that the important differences in the ratios as a function of \pt between jetty and soft events  (see Fig.\ref{Fig5}) are due to a deformation of the respective spectra and not due to a genuine change in the yields.

%
% these results show that the particle composition is slightly \so dependent suggesting that identified particle production is also different in hard and soft processes. 

\begin{table}[b]
\caption{\label{tab:table1} Particle yields, ${\rm d}N/{\rm d}y$, of kaons, protons, and phi normalized to that of pions as a function of spherocity. The number in parenthesis is the statistical error in the last digit.}
\begin{ruledtabular}
\begin{tabular}{lccc}
$S_{0}$ interval &
${\rm \frac{K^{+}+K^{-}}{\pi^{+}+\pi^{-}} }$ &
${\rm \frac{p+\bar{p}}{\pi^{+}+\pi^{-}} }$ &
${\rm \frac{2\times\phi}{\pi^{+}+\pi^{-}} }$  \\
\colrule

0.0-0.1 & 0.11267 (2) & 0.07038 (1) & 0.01035 (1)  \\
0.1-0.2 & 0.11064 (1) & 0.06836 (1) & 0.00956 (1)  \\
0.2-0.3 & 0.10967 (1) & 0.06708 (1) & 0.00930 (1)  \\
0.3-0.4 & 0.10938 (1) & 0.06628 (1) & 0.00918 (0)  \\
0.4-0.5 & 0.10938 (1) & 0.06570 (1) & 0.00913 (0)  \\
0.5-0.6 & 0.10945 (1) & 0.06528 (1) & 0.00910 (0)  \\
0.6-0.7 & 0.10962 (1) & 0.06495 (1) & 0.00908 (0)  \\
0.7-0.8 & 0.10993 (1) & 0.06472 (1) & 0.00908 (1)  \\
0.8-0.9 & 0.11043 (1) & 0.06447 (1) & 0.00907 (1)  \\
0.9-1.0 & 0.11114 (6) & 0.06408 (4) & 0.00907 (3)  \\

\end{tabular}
\end{ruledtabular}
\end{table}

%Conclusions
In summary, we have demonstrated the possibilities of the event structure analysis to separate events with very different amount of MPI and hardness. Low number of MPI will refer to enhanced hard processes samples and will be distinguished by an event structure which will be pencil-like, while the other extreme represented by \st or \so close to one will represent soft events where the maximum number of MPI is reached.  We have shown that we can very easily have access  to the nMPI by choosing the appropriate interval of \so or \st. It has to be noted that the range between the extremes is not of immediate interest since there we find the average behavior normally used to tune the models. 
It is clear that disentangling the building blocks of the collision should enhance our understanding and facilitate the theoretical calculations where so far the soft and hard processes are represented but their relative importance is often a free parameter.
The  method has been applied to the case of particle ratios showing the feasibility and bringing an interesting result, namely the completely different behavior for the hard (low \so) and soft (high \so) parts. The results suggest that a similar approach would be of interest in many other analyses where accounting for the jets is sometimes cumbersome: momentum correlations (HBT); radial flow; azimuthal flow; mean \pt vs multiplicity, etc.

The authors acknowledge the very useful discussions with Paolo Bartalini. Support for this work has been received by CONACyT under the grant
numbers 103735 and 101597; and PAPIIT-UNAM under the projects:
IN105113 and IN108414.

%\nocite{*}

\nocite{*}

\bibliography{biblio}% Produces the bibliography via BibTeX.


%\begin{thebibliography}{10}
%\providecommand{\eprint}[2][]{\url{#2}}
%
%\bibitem{SjostrandMPI} T. Sjostrand and M. Jijl, Phys. Rev. D 36, 2019 (1987). 
%
%\bibitem{banfi2004} 
%A.~Banfi~et~al., J. High Energy Phys. \textbf{08} 062 (2004).  
%
%\bibitem{ortiz2013} A. Ortiz, P. Christiansen, E. Cuautle, I. Maldonado, G. Paic
%Phys.Rev.Lett. \textbf{111} 042001 (2013)
%
%\bibitem{pythia8:0} T. Sj{\"o}strand, Mrenna, and P. Skands,
%  Computer Phys. Commun. 178, 852 (2008)
%
%\bibitem{pythia8:1} R. Corke  and T. Sjostrand, JHEP  32,  1103 (2011)
%
%\bibitem{St-alice} B. Abelev, et al. (ALICE Collaboration), Eur.Phys.J. C\textbf{72} (2012) 2124

%\end{thebibliography}
\end{document}